\documentclass[%
 reprint,
 amsmath,amssymb,
 aps,
]{revtex4-2}

\usepackage{graphicx}
\usepackage{dcolumn}
\usepackage{bm}
\usepackage{braket}
\usepackage{color}
\usepackage{appendix}

\begin{document}

\preprint{APS/123-QED}

\title{
Demonstration of the excited-state search
on the D-wave quantum annealer
}

\author{Takashi Imoto}
\email{takashi.imoto@aist.go.jp}
\affiliation{Research Center for Emerging Computing Technologies, National Institute of Advanced Industrial Science and Technology (AIST), 1-1-1 Umezono, Tsukuba, Ibaraki 305-8568, Japan.}

\author{Yuki Susa}  
\email{y-susa@nec.com}
\affiliation{Secure System Platform Research Laboratories, NEC Corporation, Kawasaki, Kanagawa 211-8666, Japan}
\affiliation{NEC-AIST Quantum Technology Cooperative Research Laboratory, National Institute of Advanced Industrial Science and Technology (AIST), Tsukuba, Ibaraki 305-8568, Japan}

\author{Ryoji Miyazaki}  
\email{miyazaki-aj@nec.com}
\affiliation{Secure System Platform Research Laboratories, NEC Corporation, Kawasaki, Kanagawa 211-8666, Japan}
\affiliation{NEC-AIST Quantum Technology Cooperative Research Laboratory, National Institute of Advanced Industrial Science and Technology (AIST), Tsukuba, Ibaraki 305-8568, Japan}

\author{Tadashi Kadowaki}  
\email{tadashi.kadowaki.j3m@jp.denso.com}
\affiliation{DENSO CORPORATION, Kounan, Minato-ku, Tokyo
108-0075, Japan}
\affiliation{Research Center for Emerging Computing Technologies, National Institute of Advanced Industrial Science and Technology (AIST), 1-1-1 Umezono, Tsukuba, Ibaraki 305-8568, Japan.}

\author{Yuichiro Matsuzaki}
\email{ymatsuzaki872@g.chuo-u.ac.jp
}
\affiliation{
Department of Electrical, Electronic, and Communication Engineering, Faculty of Science and Engineering, Chuo University, 1-13-27 Kasuga, Bunkyo-ku, Tokyo 112-8551, Japan.
}

\date{\today}

\begin{abstract}
Quantum annealing is a way to prepare an eigenstate of the problem Hamiltonian. Starting from an eigenstate of a trivial Hamiltonian, we slowly change the Hamiltonian to the problem Hamiltonian, and
the system remains in the eigenstate of the Hamiltonian as long as the so-called adiabatic condition is satisfied.
By using devices provided by D-Wave Systems Inc., there were experimental demonstrations to prepare a ground state of the problem Hamiltonian.
However, up to date,
there are no demonstrations to prepare the excited state of the problem Hamiltonian with quantum annealing.
Here, we demonstrate the excited-state search by using the D-wave processor.
The key idea is to use the reverse quantum annealing with a hot start where the initial state is the excited state of the trivial Hamiltonian. 
During the reverse quantum annealing, we control not only the transverse field but also the longitudinal field and slowly change the Hamiltonian to the problem Hamiltonian so that we can obtain the desired excited state.
As an example of the exited state search, we adopt a two-qubit Ising model as the problem Hamiltonian and succeed to prepare the excited state.
Also, we solve the shortest vector problem where the solution is embedded into the first excited state of the Ising Hamiltonian.
Our results pave the way for new applications of quantum annealers to use the excited states.
\end{abstract}


\maketitle


\section{Introduction}

Quantum annealing (QA) is a way to prepare a quantum system in an eigenstate of the non-trivial Hamiltonian, which is called the problem Hamiltonian \cite{kadowaki1998quantum, farhi2000quantum, farhi2001quantum}. The initial Hamiltonian, called the driving Hamiltonian, is chosen to be a trivial one, and we set the initial state as the eigenstate of the Hamiltonian. By slowly changing the Hamiltonian to the problem Hamiltonian, we can prepare the eigenstate of the problem Hamiltonian as long as the so-called adiabatic condition is satisfied.
One of the main applications of QA is to solve the combinational optimization problem \cite{kadowaki1998quantum, farhi2000quantum, farhi2001quantum}. The solution of the combinational optimization problem is embedded into a ground state of an Ising Hamiltonian \cite{lucas2014ising, lechner2015quantum}, and QA provides such a ground state.
D-Wave Systems Inc. developed a quantum annealer to solve such combinational optimization problems \cite{johnson2011quantum}.
 The D-Wave quantum annealing machine has been used in a variety of applications such as a quantum simulator, machine learning, and an attack on the cryptography\cite{king2018observation, kairys2020simulating, harris2018phase, zhou2021experimental, joseph2020not, joseph2021two}.
Also, the method to emulate QA using the noisy intermediate-scale quantum device(NISQ) was proposed and demonstrated \cite{li2017efficient,chen2020demonstration}.


Quantum annealing can be used for quantum chemistry \cite{aspuru2005simulated}. We can map the Hamiltonian of molecules into the Ising Hamiltonian, and the ground state of the Hamiltonian provides information about molecules, such as the prediction of the chemical reaction.
Also, there was an experimental demonstration to use the quantum annealer for the ground-state search in quantum chemistry\cite{streif2019solving}.
Also, there is a theoretical proposal to prepare the excited state of the problem Hamiltonian by using the quantum annealer\cite{seki2021excited}.
In this previous method, it is necessary to resolve the degeneracy of the driving Hamiltonian by applying inhomogeneous transverse fields. 
An initial state is prepared in the desired excited state of the driving Hamiltonian, and the Hamiltonian slowly changes into the problem Hamiltonian.
As long as the adiabatic condition is satisfied and the energy relaxation is negligible, we can obtain the desired excited state after the time evolution.
The main difficulty is to prepare the excited states of the driving Hamiltonian which is chosen as the transverse magnetic fields in the previous method.
It is known that we can use the excited state search to solve
the shortest vector problem, which is one of the post-quantum cryptography\cite{joseph2021two,ura2022analysis}.
Furthermore, the excited state search has numerous applications
in 
quantum chemistry\cite{serrano2005quantum},
quantum simulations\cite{lacroix2011introduction, tokura2020magnetic,chen2020demonstration}, and machine learning\cite{advani2020high}.
However, so far, there is no experimental demonstration of the excited-state search.

\begin{figure}[h!]
    \centering
    \includegraphics[width=90mm]{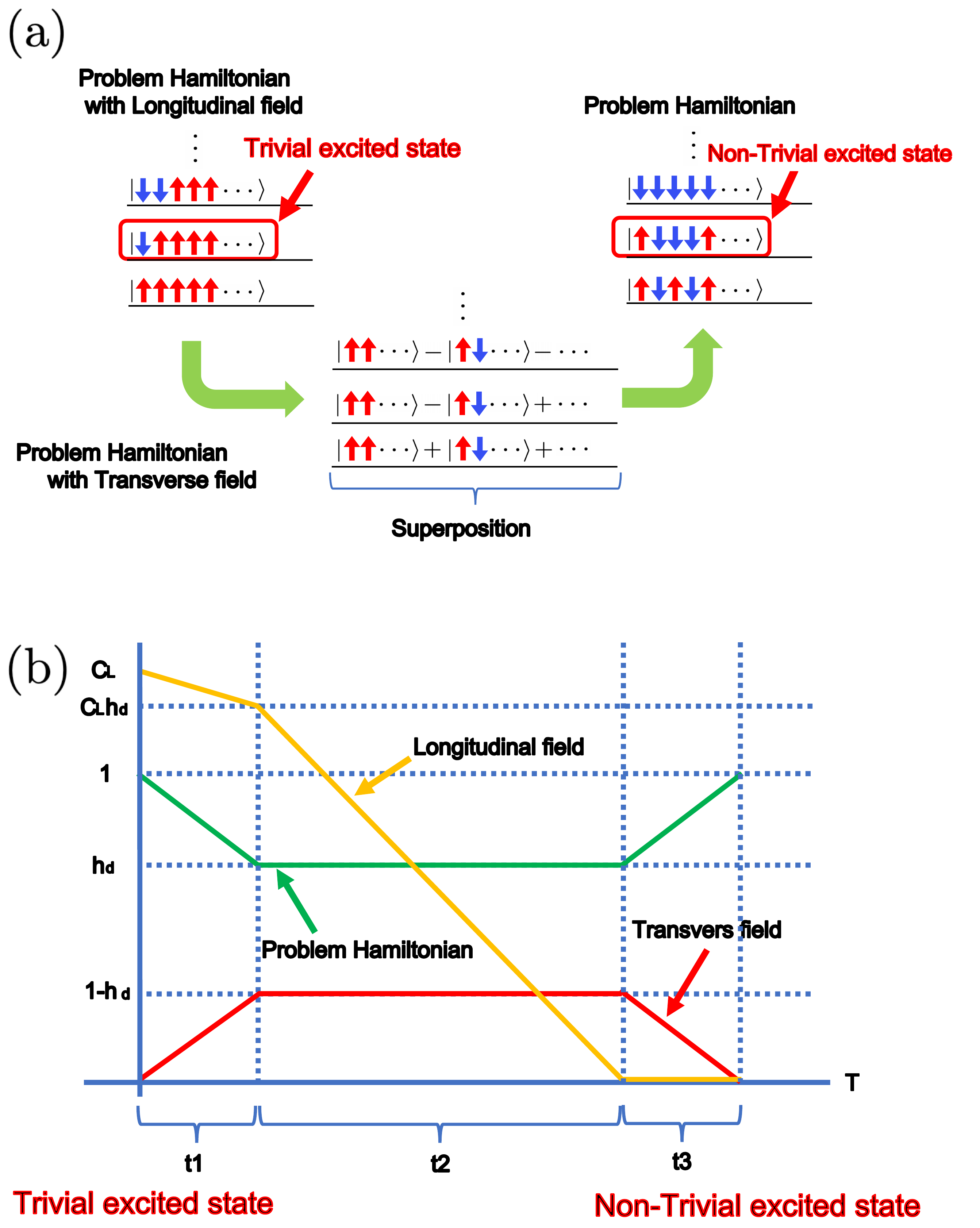}
    \caption{
    (a) We illustrate the schematic diagram of the excited state search using QA.
    First, we consider a Hamiltonian with a large longitudinal field where we can easily specify the excited state.
    We prepare the excited state in this Hamiltonian and change the Hamiltonian to the problem Hamiltonian with transverse fields. After that, we change the Hamiltonian to the problem Hamiltonian without transverse fields.
    (b) We illustrate the scheduling function of the transverse field, the longitudinal field, and the problem Hamiltonian.
    First, we decrease the problem Hamiltonian while we increase the transverse magnetic fields
    for a time period of $0 \leq t \leq t_{1}$.
    Second, we gradually turn off the longitudinal magnetic fields for a time period of
    $t_{1} \leq t \leq t_{1}+t_{2}$.
    Finally, we gradually turn off the transverse field with a time period of $t_{1}+t_{2} \leq t \leq t_{1}+t_{2}+t_{3}$.
    }
    \label{fig:concept1}
\end{figure}

In this paper, we demonstrate to prepare the excited state of the problem Hamiltonian by using the D-wave processor (see FIG. \ref{fig:concept1} (a)). 
We modify the previous method to be applicable to the current D-wave machine.
We start from a simple Hamiltonian with a large longitudinal magnetic field without transverse magnetic fields
where the first excited state can be easily inferred.
We prepare the first excited state of the
simple
Hamiltonian, and perform the reverse quantum annealing (RQA).
We control the longitudinal magnetic field during RQA, and we slowly change the Hamiltonian to the problem Hamiltonian.
To show the effectiveness of our method, we adopt a two-qubit Ising model with an inhomogeneous longitudinal magnetic field as the problem Hamiltonian and prepare the first excited state of this Hamiltonian by the D-wave processor.
Also, we apply our method to solve the shortest vector problem (SVP), which is a basis for post-quantum cryptographic protocol. 
Here, the solution of the SVP is embedded into the first excited state of the Ising Hamiltonian, and we prepare the first excited state of such an Ising Hamiltonian by using the D-wave processor.

\section{Method}

Our scheme consists of the following three steps (see FIG\ref{fig:concept1}. (a)).
First,
we start with a simple Hamiltonian where we apply a large control
longitudinal magnetic field to the problem Hamiltonian. 
In this case, we can easily calculate the first excited state (See appendix \ref{sec:make_hl}), which is the initial state of our method.
Second, during the time from 0 to $t_1$, we increase the transverse field from $0$ to $1-h_d$ while we decrease the amplitude of the problem Hamiltonian.
Third, we fix the 
the amplitude of the transverse field and turn off the control
longitudinal magnetic field during the time from $t_1$ to $t_1+t_2$. 
Finally, during the time from $t_1+t_2$ to $t_1+t_2+t_3$, we adiabatically turn off the transverse magnetic field while we increase the amplitude of the problem Hamiltonian.

The Hamiltonian is given as follows:
\begin{align}                   
    H(k)&=A(k)H_{D}+B(k)\biggl(g(k)H_{L}+H_{P}\biggr)\label{eq:gen_ann_ham}\\
    H_{D}&=-\Gamma\sum_{j=1}^{N}\sigma_{j}^{(x)}\\
    H_{L}&=\sum_{j=1}^{N}h_{j}\sigma_{j}^{(x)}.
\end{align}
where $H_{D}$ denotes the driving Hamiltonian, $H_{P}$ denotes the problem Hamiltonian, $H_{L}$ denotes the Hamiltonian of the control longitudinal magnetic field, $\Gamma$ denotes the amplitude of the transversal field, and $h_j$ denotes the amplitude of the control longitudinal field.
Also, $A(k)$, $B(k)$ and, $g(k)$ are scheduling functions defined
as follows.


\begin{align}   
    A(k)&=
    \begin{cases}
      \frac{1-h_d}{t_1}k&(0\leq k \leq t_1)\\
      1-h_d&(t_1\leq k \leq t_1+t_2)\\
      -\frac{1-h_d}{t_3}k+(1-h_d)&\!\!\!\!\!\!\frac{t_1+t_2+t_3}{t_3}\\
      &(t_1+t_2 \leq k\leq t_1+t_2+t_3)
   \end{cases}\\
   B(k)&=1-A(k)\\
   g(k)&=
      \begin{cases}
      C_{L}\ &(0\leq k \leq t_1)\\
      -\frac{C_{L}}{t_2}(k-t_1)+C_{L}\ &(t_1\leq k \leq t_1+t_2)\\
      0\ &(t_1+t_2\leq k \leq t_1+t_2+t_3)
   \end{cases}
\end{align}
where $1-h_d$ denotes the maximum amplitude of the transverse magnetic field in our algorithms and $C_L$ denotes the initial amplitude of the longitudinal magnetic field.
The illustration of the scheduling function is given in FIG\ref{fig:concept1}. (b).

\section{Excited-state search of the
two-qubit Ising model}
\label{sec:two_qubit_case}

\begin{figure*}[t!]
\includegraphics[width=185mm]{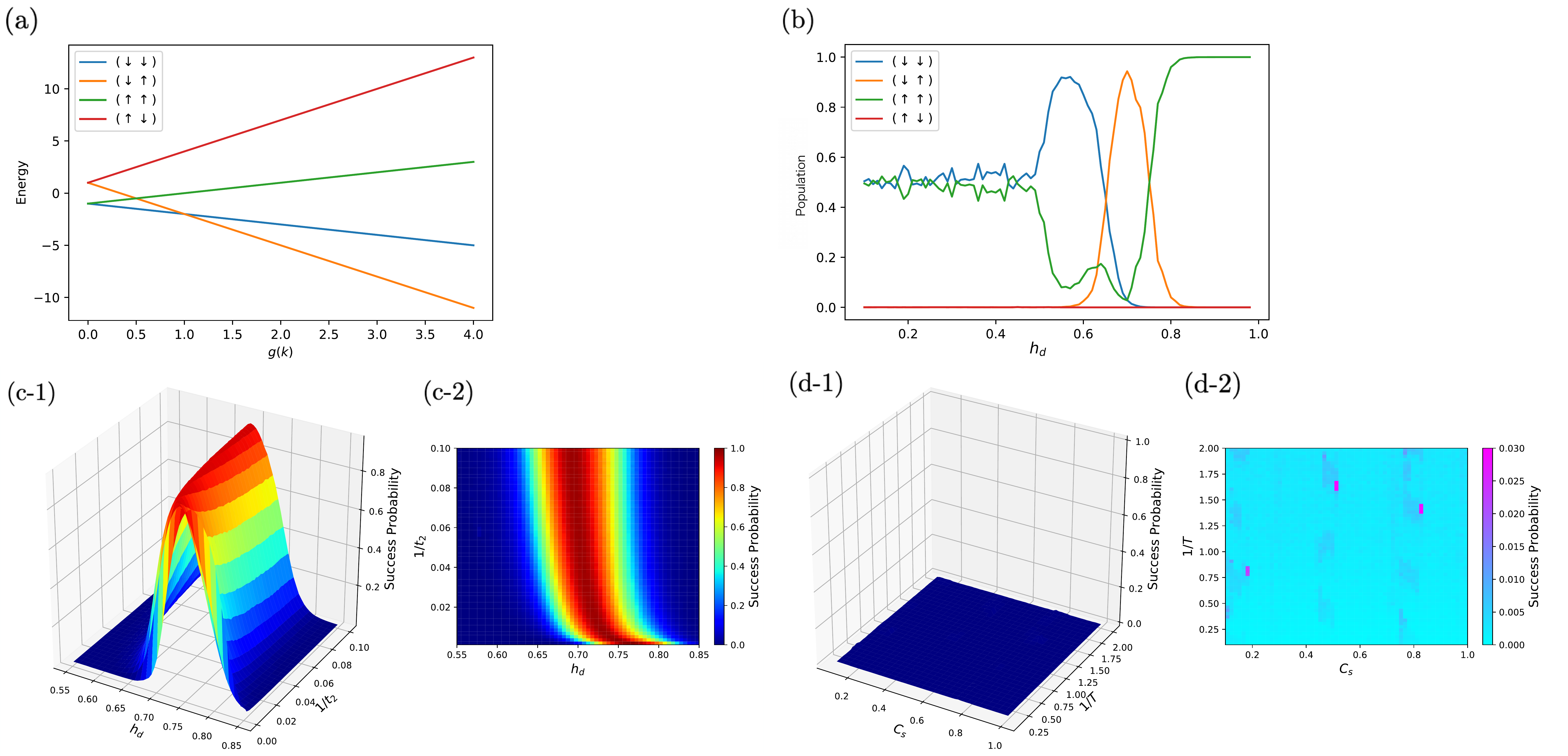}
\caption{
(a)
The energy diagram against $g(k)$ which determines the ratio between the problem Hamiltonian and the Hamiltonian with a longitudinal magnetic field. 
If the excited-state search succeeds, we should obtain $\ket{\uparrow\downarrow}$ or $\ket{\downarrow\uparrow}$ as the final state, while the initial state is $\ket{\uparrow\uparrow}$ as the second excited state of the problem Hamiltonian.
(b)
The plot of the population of the computational basis states against the strength of the transverse field.
Here, we set $t_1=2$, $t_2=20$, and, $t_3=2$ and the shot number is $100000$.
Between $0.8\leq h_d\leq 1.0$, the population of $\ket{\uparrow\downarrow}$, which is the same as the initial state, is dominant.
This comes from the fact that, for weaker transverse magnetic fields, the freezing effect becomes stronger.
On the other hand, between $0.0\leq h_d\leq 0.6$, the population of $\ket{\downarrow\uparrow}$, which is the ground state of the problem Hamiltonian is dominant.
This is because the system relaxes into the ground state due to the strong decoherence.
We find that, for $h_d\simeq 0.7$, we can maximize the success probability which is the sum of the populations of $\ket{\uparrow\downarrow}$ and $\ket{\downarrow\uparrow}$.
(c-1), (c-2)
The success probability of our method against the $h_d$ and $t_2$ when the shot number is $100000$.
As we decrease $h_d$, the optimized $t_2$ becomes smaller.
(d-1), (d-2)
The success probability of the conventional QA for the excited-state search against the $h_d$ and $t_2$ in  $100000$ shot number.
The success probability of the conventional method is much lower than that of ours.
}\label{fig:D-wave_result_two_qubit}
\end{figure*}


To demonstrate our method by using the D-wave processor, we adopt a two-qubit Ising model 
as the problem Hamiltonian while we adopt
an inhomogeneous longitudinal magnetic field 
as 
the control Hamiltonian. 
In this case, $H_L$ and $H_P$ are given by
\begin{align}
    H_{L}&=2\hat{\sigma}^{z}_{1}-\hat{\sigma}^{z}_{2}\\
    H_{P}&=-J\hat{\sigma}^{z}_{1}\hat{\sigma}^{z}_{2}.
\end{align}
Here, we set $J=1$, $C_{L}=4$, $t_1=2$, $t_2=20$ and, $t_3=2$.
Also, the initial state is
$\ket{\uparrow\uparrow}$.
From FIG. \ref{fig:D-wave_result_two_qubit} (a), we plot the energy diagram against $g(k)$, and the final states of a successful excited state search are $\ket{\uparrow\downarrow}$ and $\ket{\downarrow\uparrow}$.

Let us discuss how the values of $h_d$ affect the performance of our method.
For $h_d\simeq 1$, only a weak transverse magnetic field is applied during the RQA, and the system remains in the initial state. We call these phenomena "freezing".
On the other hand, for $h_d\simeq 0$, we apply a strong transverse magnetic field, and the energy relaxation time becomes small \cite{imoto2023measurement}, which makes it difficult for the system to remain in the excited state.
For these reasons, it is important for our method to optimize the value of $h_d$.

FIG \ref{fig:D-wave_result_two_qubit} (b) shows that we successfully obtain the target excited state for $h_d\simeq 0.7$.
On the other hand, for $h_d>0.8$, 
the freezing occurs, and
the dominant state after performing our protocol is $\ket{\uparrow\uparrow}$, which is the same as the initial state. For $h_d<0.6$, an energy relaxation is relevant, and so the dominant states after performing our protocol are $\ket{\uparrow\uparrow}$ and $\ket{\downarrow\downarrow}$, which are the ground states of the problem Hamiltonian.


In FIG \ref{fig:D-wave_result_two_qubit} (c-1) and (c-2),
we illustrate that the success probability depends not only on $h_d$ but also on $t_2$.
As we decrease $h_d$, the optimized $t_2$ decreases.
We can understand this as follows.
For a smaller $h_d$, we apply more transverse magnetic fields, and so we can suppress the freezing while the energy relaxation becomes larger.
In this case, for the excited state search, we need to decrease $t_2$ so that the system could remain in the excited state.

For comparison,
we also
plot the population of the target excited state 
using the non-adiabatic transition of the conventional QA. 
We compare the success probability 
of this conventional method
with that of our method. For the conventional QA, the annealing Hamiltonian is described as follows:
\begin{align}
    H(t)=\biggl(1-\frac{t}{T}\biggr)H_{D}+\frac{t}{T}C_{s}H_{P}\label{eq:conv_ann_ham}
\end{align}
where $T$ denotes the annealing time and $C_{s}$ denotes the amplitude of the problem Hamiltonian.
We prepare the ground state of $H_{D}$ at $t=0$, and let the state evolve by the Hamiltonian.
In the conventional QA, if non-adiabatic transitions occur, we could obtain a finite population of the excited state by optimizing $T$ and $C_s$.
From FIG \ref{fig:D-wave_result_two_qubit}, 
we confirm that, as we decrease $T$, the success probability to obtain the desired excited state becomes larger.
This is because the non-adiabatic transitions become more relevant for a shorter $T$.
However, the success probability with optimized parameters with the conventional QA is much smaller than that of our method. This shows 
that, for the excited-state search, our method is superior to the conventional method.

\section{Solving the
shortest-vector problem
by the excited-state search
}

\begin{figure*}[t!]
\includegraphics[width=180mm]{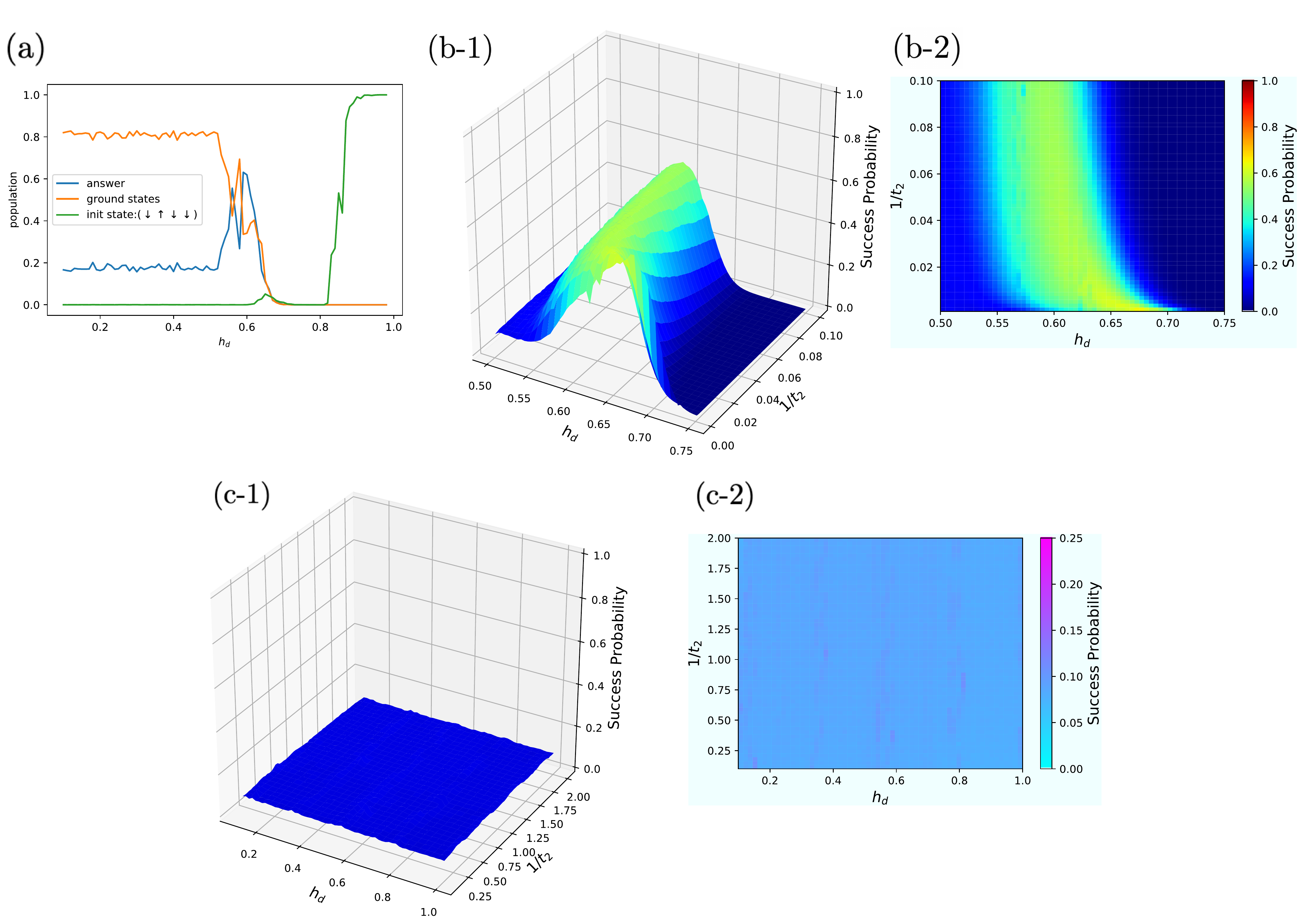}
\caption{(a) The populations of each state which is written on a computational basis against the parameter concerning the strength of the transverse field.
In this figure, we set the annealing time $t_1=2$, $t_2=20$, and, $t_3=2$ and the shot number is $1000$.
It can be seen that the population of
the state $\ket{\downarrow\uparrow\downarrow\downarrow}$, which is the same as the initial state, is dominant between $0.8\leq h_d\leq 1.0$.
This is due to the weak transverse magnetic field and the strong effect of freezing.
On the other hand, the ratio of the ground states of the problem Hamiltonian is dominant between $0.0\leq h_d\leq 0.5$.
This is because the effect of decoherence becomes stronger, causing a transition to the ground state.
When $h_d$ is around $0.6$, the population of the desired excited state becomes dominant and the excited state search is successful.
(b-1), (b-2) The success probability of the SVP against the $h_d$ and $t_2$ with our method when the shot number is $100000$.
As we increase $h_d$, the optimized $t_2$ becomes larger. 
(c-1), (c-2)
The success probability of the SVP against $h_d$ and $t_2$ with the conventional method in  $100000$ shot number.
Since we need to search the excited state for the SVP, it is difficult to obtain a high success probability with
the conventional method, which was originally developed for the ground-state search.
}\label{fig:D-wave_result_three_qubit}
\end{figure*}


We apply the excited state search to solving the SVP.
This problem plays a central role in Lattice-Based Cryptography, which is a potential candidate for post-quantum cryptography.
It is known that the excited state search is useful to solve this problem\cite{ura2022analysis}.
The SVP is the problem to find the shortest non-zero vector in a given lattice where the number of bases is $N$.
Among several ways to transform the SVP into the Ising Hamiltonian, we adopt the Hamming encoding.
We set $N=2$, and we use 4 qubits to describe the problem Hamiltonian.
We show the details about how to map the SVP to the Ising Hamiltonian in the appendix \ref{sec:mapping}.
The problem Hamiltonian and the control longitudinal field are given by

\begin{align}
    H_{L}&=\frac{1}{2}h_{1}\hat{\sigma}^{z}_{1}
    +\frac{1}{2}h_{2}\hat{\sigma}^{z}_{2}
    +\frac{1}{2}h_{3}\hat{\sigma}^{z}_{3}
    +\frac{1}{2}h_{4}\hat{\sigma}^{z}_{4}\\
    H_{P}&=\frac{1}{2}G_{11}\hat{\sigma}^{z}_{1}\hat{\sigma}^{z}_{2}
    +\frac{1}{2}G_{22}\hat{\sigma}^{z}_{3}\hat{\sigma}^{z}_{4}
    +\frac{1}{2}G_{12}\hat{\sigma}^{z}_{1}\hat{\sigma}^{z}_{3}
    +\frac{1}{2}G_{12}\hat{\sigma}^{z}_{1}\hat{\sigma}^{z}_{4}\notag\\
    &\ \ +\frac{1}{2}G_{12}\hat{\sigma}^{z}_{2}\hat{\sigma}^{z}_{3}+\frac{1}{2}G_{12}\hat{\sigma}^{z}_{2}\hat{\sigma}^{z}_{4}
\end{align}
where $G_{ij}\equiv\vec{b}_{i}\cdot\vec{b}_j$ denotes the Gram matrix and $\{\vec{b}_{j}\}_{j=1}^{2}$ denotes a set of linearly independent vectors determined by the problem.
We set $|\vec{b}_1|=1.0$ and $\vec{b}_2=1.3$. Also, the angle between the $\vec{b}_1$ and $\vec{b}_2$ is set to be $\theta=\pi/10$. 
Furthermore, we set $h_{1}=4$, $h_{2}=4$, $h_{3}=1$, $h_{4}=2$, and $C_{L}=4$.
It is worth mentioning that the ground states and the first excited states of the problem Hamiltonian are degenerate.
So, in order to solve the SVP, we set the initial state as $\ket{\downarrow\uparrow\downarrow\downarrow}$, which is the 2-th excited state of the initial Hamiltonian
(see Appendix \ref{sec:mapping}).

We plot the result in FIG. \ref{fig:D-wave_result_three_qubit} (a), (b-1) and, (b-2).
FIG. \ref{fig:D-wave_result_three_qubit}(a) shows that the population of $\ket{\uparrow\uparrow\downarrow\downarrow}$, which is one of the degenerate first excited states of the problem Hamiltonian, is dominant around $h_{d}=0.6$.
On the other hand, the population of the other first excited state $\ket{\downarrow\downarrow\uparrow\uparrow}$ is much smaller.
The reason for this is that the first excited state, which is close to the initial state, is preferentially obtained.
In regions where $h_d$ is more than $0.8$, the population of the initial state $\ket{\downarrow\uparrow\downarrow\downarrow}$ is dominant. 
This is due to freezing.
In the region where $h_d$ is less than $0.5$, the population of the ground states is dominant due to the energy relaxation.
FIG \ref{fig:D-wave_result_three_qubit} (b-1) and (b-2) illustrate how the success probability depends on $h_d$ and $t_2$.
Similar to the two-qubit case in the section \ref{sec:two_qubit_case}, the optimized $t_2$ becomes larger as we increase $h_d$.
FIG \ref{fig:D-wave_result_three_qubit} (c-1) and (c-2) illustrated the success probability of the SVP with the conventional method.
In this case, we use the Eq (\ref{eq:conv_ann_ham}) as the annealing Hamiltonian.
The success probability of the conventional method, which was originally developed for the ground-state search, is much smaller than that of our method.
This is because the solution is encoded in the excited state for the SVP.



\section{Conclusion}
In conclusion, we proposed and demonstrated a method to prepare an excited state of the problem Hamiltonian by using a D-wave processor.
We use the reverse quantum annealing with a hot start.
More specifically,
we start from the first excited state of a trivial Ising Hamiltonian and slowly change the Hamiltonian to the problem Hamiltonian.
During the reverse quantum annealing,
we control both the transverse field
and the longitudinal field.
We adopt a two-qubit Ising model as the problem Hamiltonian, and we succeed to prepare the excited state.
Moreover, by using our method, we solve the shortest vector problem where the solution is embedded into the first excited state of the Ising Hamiltonian.
Our results pave the way for new applications of quantum annealers to use the excited states.

This paper is partly
based on results obtained from a project, JPNP16007,
commissioned by the New Energy and Industrial Technology Development Organization (NEDO), Japan. This
work was supported by JST Moonshot R\&D (Grant
Number JPMJMS226C).
YM is supported by JSPS KAKENHI (Grant Number 23H04390).


\nocite{*}

\bibliography{apssamp}

\providecommand{\noopsort}[1]{}\providecommand{\singleletter}[1]{#1}%
\begin{thebibliography}{36}%
\makeatletter
\providecommand \@ifxundefined [1]{%
 \@ifx{#1\undefined}
}%
\providecommand \@ifnum [1]{%
 \ifnum #1\expandafter \@firstoftwo
 \else \expandafter \@secondoftwo
 \fi
}%
\providecommand \@ifx [1]{%
 \ifx #1\expandafter \@firstoftwo
 \else \expandafter \@secondoftwo
 \fi
}%
\providecommand \natexlab [1]{#1}%
\providecommand \enquote  [1]{``#1''}%
\providecommand \bibnamefont  [1]{#1}%
\providecommand \bibfnamefont [1]{#1}%
\providecommand \citenamefont [1]{#1}%
\providecommand \href@noop [0]{\@secondoftwo}%
\providecommand \href [0]{\begingroup \@sanitize@url \@href}%
\providecommand \@href[1]{\@@startlink{#1}\@@href}%
\providecommand \@@href[1]{\endgroup#1\@@endlink}%
\providecommand \@sanitize@url [0]{\catcode `\\12\catcode `\$12\catcode
  `\&12\catcode `\#12\catcode `\^12\catcode `\_12\catcode `\%12\relax}%
\providecommand \@@startlink[1]{}%
\providecommand \@@endlink[0]{}%
\providecommand \url  [0]{\begingroup\@sanitize@url \@url }%
\providecommand \@url [1]{\endgroup\@href {#1}{\urlprefix }}%
\providecommand \urlprefix  [0]{URL }%
\providecommand \Eprint [0]{\href }%
\providecommand \doibase [0]{https://doi.org/}%
\providecommand \selectlanguage [0]{\@gobble}%
\providecommand \bibinfo  [0]{\@secondoftwo}%
\providecommand \bibfield  [0]{\@secondoftwo}%
\providecommand \translation [1]{[#1]}%
\providecommand \BibitemOpen [0]{}%
\providecommand \bibitemStop [0]{}%
\providecommand \bibitemNoStop [0]{.\EOS\space}%
\providecommand \EOS [0]{\spacefactor3000\relax}%
\providecommand \BibitemShut  [1]{\csname bibitem#1\endcsname}%
\let\auto@bib@innerbib\@empty
\bibitem [{\citenamefont {Kadowaki}\ and\ \citenamefont
  {Nishimori}(1998)}]{kadowaki1998quantum}%
  \BibitemOpen
  \bibfield  {author} {\bibinfo {author} {\bibfnamefont {T.}~\bibnamefont
  {Kadowaki}}\ and\ \bibinfo {author} {\bibfnamefont {H.}~\bibnamefont
  {Nishimori}},\ }\bibfield  {title} {\bibinfo {title} {Quantum annealing in
  the transverse ising model},\ }\href@noop {} {\bibfield  {journal} {\bibinfo
  {journal} {Physical Review E}\ }\textbf {\bibinfo {volume} {58}},\ \bibinfo
  {pages} {5355} (\bibinfo {year} {1998})}\BibitemShut {NoStop}%
\bibitem [{\citenamefont {Farhi}\ \emph {et~al.}(2000)\citenamefont {Farhi},
  \citenamefont {Goldstone}, \citenamefont {Gutmann},\ and\ \citenamefont
  {Sipser}}]{farhi2000quantum}%
  \BibitemOpen
  \bibfield  {author} {\bibinfo {author} {\bibfnamefont {E.}~\bibnamefont
  {Farhi}}, \bibinfo {author} {\bibfnamefont {J.}~\bibnamefont {Goldstone}},
  \bibinfo {author} {\bibfnamefont {S.}~\bibnamefont {Gutmann}},\ and\ \bibinfo
  {author} {\bibfnamefont {M.}~\bibnamefont {Sipser}},\ }\bibfield  {title}
  {\bibinfo {title} {Quantum computation by adiabatic evolution},\ }\href@noop
  {} {\bibfield  {journal} {\bibinfo  {journal} {arXiv preprint
  quant-ph/0001106}\ } (\bibinfo {year} {2000})}\BibitemShut {NoStop}%
\bibitem [{\citenamefont {Farhi}\ \emph {et~al.}(2001)\citenamefont {Farhi},
  \citenamefont {Goldstone}, \citenamefont {Gutmann}, \citenamefont {Lapan},
  \citenamefont {Lundgren},\ and\ \citenamefont {Preda}}]{farhi2001quantum}%
  \BibitemOpen
  \bibfield  {author} {\bibinfo {author} {\bibfnamefont {E.}~\bibnamefont
  {Farhi}}, \bibinfo {author} {\bibfnamefont {J.}~\bibnamefont {Goldstone}},
  \bibinfo {author} {\bibfnamefont {S.}~\bibnamefont {Gutmann}}, \bibinfo
  {author} {\bibfnamefont {J.}~\bibnamefont {Lapan}}, \bibinfo {author}
  {\bibfnamefont {A.}~\bibnamefont {Lundgren}},\ and\ \bibinfo {author}
  {\bibfnamefont {D.}~\bibnamefont {Preda}},\ }\bibfield  {title} {\bibinfo
  {title} {A quantum adiabatic evolution algorithm applied to random instances
  of an np-complete problem},\ }\href@noop {} {\bibfield  {journal} {\bibinfo
  {journal} {Science}\ }\textbf {\bibinfo {volume} {292}},\ \bibinfo {pages}
  {472} (\bibinfo {year} {2001})}\BibitemShut {NoStop}%
\bibitem [{\citenamefont {Lucas}(2014)}]{lucas2014ising}%
  \BibitemOpen
  \bibfield  {author} {\bibinfo {author} {\bibfnamefont {A.}~\bibnamefont
  {Lucas}},\ }\bibfield  {title} {\bibinfo {title} {Ising formulations of many
  np problems},\ }\href@noop {} {\bibfield  {journal} {\bibinfo  {journal}
  {Frontiers in physics}\ }\textbf {\bibinfo {volume} {2}},\ \bibinfo {pages}
  {5} (\bibinfo {year} {2014})}\BibitemShut {NoStop}%
\bibitem [{\citenamefont {Lechner}\ \emph {et~al.}(2015)\citenamefont
  {Lechner}, \citenamefont {Hauke},\ and\ \citenamefont
  {Zoller}}]{lechner2015quantum}%
  \BibitemOpen
  \bibfield  {author} {\bibinfo {author} {\bibfnamefont {W.}~\bibnamefont
  {Lechner}}, \bibinfo {author} {\bibfnamefont {P.}~\bibnamefont {Hauke}},\
  and\ \bibinfo {author} {\bibfnamefont {P.}~\bibnamefont {Zoller}},\
  }\bibfield  {title} {\bibinfo {title} {A quantum annealing architecture with
  all-to-all connectivity from local interactions},\ }\href@noop {} {\bibfield
  {journal} {\bibinfo  {journal} {Science advances}\ }\textbf {\bibinfo
  {volume} {1}},\ \bibinfo {pages} {e1500838} (\bibinfo {year}
  {2015})}\BibitemShut {NoStop}%
\bibitem [{\citenamefont {Johnson}\ \emph {et~al.}(2011)\citenamefont
  {Johnson}, \citenamefont {Amin}, \citenamefont {Gildert}, \citenamefont
  {Lanting}, \citenamefont {Hamze}, \citenamefont {Dickson}, \citenamefont
  {Harris}, \citenamefont {Berkley}, \citenamefont {Johansson}, \citenamefont
  {Bunyk} \emph {et~al.}}]{johnson2011quantum}%
  \BibitemOpen
  \bibfield  {author} {\bibinfo {author} {\bibfnamefont {M.~W.}\ \bibnamefont
  {Johnson}}, \bibinfo {author} {\bibfnamefont {M.~H.}\ \bibnamefont {Amin}},
  \bibinfo {author} {\bibfnamefont {S.}~\bibnamefont {Gildert}}, \bibinfo
  {author} {\bibfnamefont {T.}~\bibnamefont {Lanting}}, \bibinfo {author}
  {\bibfnamefont {F.}~\bibnamefont {Hamze}}, \bibinfo {author} {\bibfnamefont
  {N.}~\bibnamefont {Dickson}}, \bibinfo {author} {\bibfnamefont
  {R.}~\bibnamefont {Harris}}, \bibinfo {author} {\bibfnamefont {A.~J.}\
  \bibnamefont {Berkley}}, \bibinfo {author} {\bibfnamefont {J.}~\bibnamefont
  {Johansson}}, \bibinfo {author} {\bibfnamefont {P.}~\bibnamefont {Bunyk}},
  \emph {et~al.},\ }\bibfield  {title} {\bibinfo {title} {Quantum annealing
  with manufactured spins},\ }\href@noop {} {\bibfield  {journal} {\bibinfo
  {journal} {Nature}\ }\textbf {\bibinfo {volume} {473}},\ \bibinfo {pages}
  {194} (\bibinfo {year} {2011})}\BibitemShut {NoStop}%
\bibitem [{\citenamefont {King}\ \emph {et~al.}(2018)\citenamefont {King},
  \citenamefont {Carrasquilla}, \citenamefont {Raymond}, \citenamefont
  {Ozfidan}, \citenamefont {Andriyash}, \citenamefont {Berkley}, \citenamefont
  {Reis}, \citenamefont {Lanting}, \citenamefont {Harris}, \citenamefont
  {Altomare} \emph {et~al.}}]{king2018observation}%
  \BibitemOpen
  \bibfield  {author} {\bibinfo {author} {\bibfnamefont {A.~D.}\ \bibnamefont
  {King}}, \bibinfo {author} {\bibfnamefont {J.}~\bibnamefont {Carrasquilla}},
  \bibinfo {author} {\bibfnamefont {J.}~\bibnamefont {Raymond}}, \bibinfo
  {author} {\bibfnamefont {I.}~\bibnamefont {Ozfidan}}, \bibinfo {author}
  {\bibfnamefont {E.}~\bibnamefont {Andriyash}}, \bibinfo {author}
  {\bibfnamefont {A.}~\bibnamefont {Berkley}}, \bibinfo {author} {\bibfnamefont
  {M.}~\bibnamefont {Reis}}, \bibinfo {author} {\bibfnamefont {T.}~\bibnamefont
  {Lanting}}, \bibinfo {author} {\bibfnamefont {R.}~\bibnamefont {Harris}},
  \bibinfo {author} {\bibfnamefont {F.}~\bibnamefont {Altomare}}, \emph
  {et~al.},\ }\bibfield  {title} {\bibinfo {title} {Observation of topological
  phenomena in a programmable lattice of 1,800 qubits},\ }\href@noop {}
  {\bibfield  {journal} {\bibinfo  {journal} {Nature}\ }\textbf {\bibinfo
  {volume} {560}},\ \bibinfo {pages} {456} (\bibinfo {year}
  {2018})}\BibitemShut {NoStop}%
\bibitem [{\citenamefont {Kairys}\ \emph {et~al.}(2020)\citenamefont {Kairys},
  \citenamefont {King}, \citenamefont {Ozfidan}, \citenamefont {Boothby},
  \citenamefont {Raymond}, \citenamefont {Banerjee},\ and\ \citenamefont
  {Humble}}]{kairys2020simulating}%
  \BibitemOpen
  \bibfield  {author} {\bibinfo {author} {\bibfnamefont {P.}~\bibnamefont
  {Kairys}}, \bibinfo {author} {\bibfnamefont {A.~D.}\ \bibnamefont {King}},
  \bibinfo {author} {\bibfnamefont {I.}~\bibnamefont {Ozfidan}}, \bibinfo
  {author} {\bibfnamefont {K.}~\bibnamefont {Boothby}}, \bibinfo {author}
  {\bibfnamefont {J.}~\bibnamefont {Raymond}}, \bibinfo {author} {\bibfnamefont
  {A.}~\bibnamefont {Banerjee}},\ and\ \bibinfo {author} {\bibfnamefont
  {T.~S.}\ \bibnamefont {Humble}},\ }\bibfield  {title} {\bibinfo {title}
  {Simulating the shastry-sutherland ising model using quantum annealing},\
  }\href@noop {} {\bibfield  {journal} {\bibinfo  {journal} {Prx Quantum}\
  }\textbf {\bibinfo {volume} {1}},\ \bibinfo {pages} {020320} (\bibinfo {year}
  {2020})}\BibitemShut {NoStop}%
\bibitem [{\citenamefont {Harris}\ \emph {et~al.}(2018)\citenamefont {Harris},
  \citenamefont {Sato}, \citenamefont {Berkley}, \citenamefont {Reis},
  \citenamefont {Altomare}, \citenamefont {Amin}, \citenamefont {Boothby},
  \citenamefont {Bunyk}, \citenamefont {Deng}, \citenamefont {Enderud} \emph
  {et~al.}}]{harris2018phase}%
  \BibitemOpen
  \bibfield  {author} {\bibinfo {author} {\bibfnamefont {R.}~\bibnamefont
  {Harris}}, \bibinfo {author} {\bibfnamefont {Y.}~\bibnamefont {Sato}},
  \bibinfo {author} {\bibfnamefont {A.}~\bibnamefont {Berkley}}, \bibinfo
  {author} {\bibfnamefont {M.}~\bibnamefont {Reis}}, \bibinfo {author}
  {\bibfnamefont {F.}~\bibnamefont {Altomare}}, \bibinfo {author}
  {\bibfnamefont {M.}~\bibnamefont {Amin}}, \bibinfo {author} {\bibfnamefont
  {K.}~\bibnamefont {Boothby}}, \bibinfo {author} {\bibfnamefont
  {P.}~\bibnamefont {Bunyk}}, \bibinfo {author} {\bibfnamefont
  {C.}~\bibnamefont {Deng}}, \bibinfo {author} {\bibfnamefont {C.}~\bibnamefont
  {Enderud}}, \emph {et~al.},\ }\bibfield  {title} {\bibinfo {title} {Phase
  transitions in a programmable quantum spin glass simulator},\ }\href@noop {}
  {\bibfield  {journal} {\bibinfo  {journal} {Science}\ }\textbf {\bibinfo
  {volume} {361}},\ \bibinfo {pages} {162} (\bibinfo {year}
  {2018})}\BibitemShut {NoStop}%
\bibitem [{\citenamefont {Zhou}\ \emph {et~al.}(2021)\citenamefont {Zhou},
  \citenamefont {Green}, \citenamefont {Dahl},\ and\ \citenamefont
  {Chamon}}]{zhou2021experimental}%
  \BibitemOpen
  \bibfield  {author} {\bibinfo {author} {\bibfnamefont {S.}~\bibnamefont
  {Zhou}}, \bibinfo {author} {\bibfnamefont {D.}~\bibnamefont {Green}},
  \bibinfo {author} {\bibfnamefont {E.~D.}\ \bibnamefont {Dahl}},\ and\
  \bibinfo {author} {\bibfnamefont {C.}~\bibnamefont {Chamon}},\ }\bibfield
  {title} {\bibinfo {title} {Experimental realization of classical z 2 spin
  liquids in a programmable quantum device},\ }\href@noop {} {\bibfield
  {journal} {\bibinfo  {journal} {Physical Review B}\ }\textbf {\bibinfo
  {volume} {104}},\ \bibinfo {pages} {L081107} (\bibinfo {year}
  {2021})}\BibitemShut {NoStop}%
\bibitem [{\citenamefont {Joseph}\ \emph {et~al.}(2020)\citenamefont {Joseph},
  \citenamefont {Ghionis}, \citenamefont {Ling},\ and\ \citenamefont
  {Mintert}}]{joseph2020not}%
  \BibitemOpen
  \bibfield  {author} {\bibinfo {author} {\bibfnamefont {D.}~\bibnamefont
  {Joseph}}, \bibinfo {author} {\bibfnamefont {A.}~\bibnamefont {Ghionis}},
  \bibinfo {author} {\bibfnamefont {C.}~\bibnamefont {Ling}},\ and\ \bibinfo
  {author} {\bibfnamefont {F.}~\bibnamefont {Mintert}},\ }\bibfield  {title}
  {\bibinfo {title} {Not-so-adiabatic quantum computation for the shortest
  vector problem},\ }\href@noop {} {\bibfield  {journal} {\bibinfo  {journal}
  {Physical Review Research}\ }\textbf {\bibinfo {volume} {2}},\ \bibinfo
  {pages} {013361} (\bibinfo {year} {2020})}\BibitemShut {NoStop}%
\bibitem [{\citenamefont {Joseph}\ \emph {et~al.}(2021)\citenamefont {Joseph},
  \citenamefont {Callison}, \citenamefont {Ling},\ and\ \citenamefont
  {Mintert}}]{joseph2021two}%
  \BibitemOpen
  \bibfield  {author} {\bibinfo {author} {\bibfnamefont {D.}~\bibnamefont
  {Joseph}}, \bibinfo {author} {\bibfnamefont {A.}~\bibnamefont {Callison}},
  \bibinfo {author} {\bibfnamefont {C.}~\bibnamefont {Ling}},\ and\ \bibinfo
  {author} {\bibfnamefont {F.}~\bibnamefont {Mintert}},\ }\bibfield  {title}
  {\bibinfo {title} {Two quantum ising algorithms for the shortest-vector
  problem},\ }\href@noop {} {\bibfield  {journal} {\bibinfo  {journal}
  {Physical Review A}\ }\textbf {\bibinfo {volume} {103}},\ \bibinfo {pages}
  {032433} (\bibinfo {year} {2021})}\BibitemShut {NoStop}%
\bibitem [{\citenamefont {Li}\ and\ \citenamefont
  {Benjamin}(2017)}]{li2017efficient}%
  \BibitemOpen
  \bibfield  {author} {\bibinfo {author} {\bibfnamefont {Y.}~\bibnamefont
  {Li}}\ and\ \bibinfo {author} {\bibfnamefont {S.~C.}\ \bibnamefont
  {Benjamin}},\ }\bibfield  {title} {\bibinfo {title} {Efficient variational
  quantum simulator incorporating active error minimization},\ }\href@noop {}
  {\bibfield  {journal} {\bibinfo  {journal} {Physical Review X}\ }\textbf
  {\bibinfo {volume} {7}},\ \bibinfo {pages} {021050} (\bibinfo {year}
  {2017})}\BibitemShut {NoStop}%
\bibitem [{\citenamefont {Chen}\ \emph {et~al.}(2020)\citenamefont {Chen},
  \citenamefont {Gong}, \citenamefont {Xu}, \citenamefont {Yuan}, \citenamefont
  {Wang}, \citenamefont {Wang}, \citenamefont {Ying}, \citenamefont {Lin},
  \citenamefont {Xu}, \citenamefont {Wu} \emph
  {et~al.}}]{chen2020demonstration}%
  \BibitemOpen
  \bibfield  {author} {\bibinfo {author} {\bibfnamefont {M.-C.}\ \bibnamefont
  {Chen}}, \bibinfo {author} {\bibfnamefont {M.}~\bibnamefont {Gong}}, \bibinfo
  {author} {\bibfnamefont {X.}~\bibnamefont {Xu}}, \bibinfo {author}
  {\bibfnamefont {X.}~\bibnamefont {Yuan}}, \bibinfo {author} {\bibfnamefont
  {J.-W.}\ \bibnamefont {Wang}}, \bibinfo {author} {\bibfnamefont
  {C.}~\bibnamefont {Wang}}, \bibinfo {author} {\bibfnamefont {C.}~\bibnamefont
  {Ying}}, \bibinfo {author} {\bibfnamefont {J.}~\bibnamefont {Lin}}, \bibinfo
  {author} {\bibfnamefont {Y.}~\bibnamefont {Xu}}, \bibinfo {author}
  {\bibfnamefont {Y.}~\bibnamefont {Wu}}, \emph {et~al.},\ }\bibfield  {title}
  {\bibinfo {title} {Demonstration of adiabatic variational quantum computing
  with a superconducting quantum coprocessor},\ }\href@noop {} {\bibfield
  {journal} {\bibinfo  {journal} {Physical Review Letters}\ }\textbf {\bibinfo
  {volume} {125}},\ \bibinfo {pages} {180501} (\bibinfo {year}
  {2020})}\BibitemShut {NoStop}%
\bibitem [{\citenamefont {Aspuru-Guzik}\ \emph {et~al.}(2005)\citenamefont
  {Aspuru-Guzik}, \citenamefont {Dutoi}, \citenamefont {Love},\ and\
  \citenamefont {Head-Gordon}}]{aspuru2005simulated}%
  \BibitemOpen
  \bibfield  {author} {\bibinfo {author} {\bibfnamefont {A.}~\bibnamefont
  {Aspuru-Guzik}}, \bibinfo {author} {\bibfnamefont {A.~D.}\ \bibnamefont
  {Dutoi}}, \bibinfo {author} {\bibfnamefont {P.~J.}\ \bibnamefont {Love}},\
  and\ \bibinfo {author} {\bibfnamefont {M.}~\bibnamefont {Head-Gordon}},\
  }\bibfield  {title} {\bibinfo {title} {Simulated quantum computation of
  molecular energies},\ }\href@noop {} {\bibfield  {journal} {\bibinfo
  {journal} {Science}\ }\textbf {\bibinfo {volume} {309}},\ \bibinfo {pages}
  {1704} (\bibinfo {year} {2005})}\BibitemShut {NoStop}%
\bibitem [{\citenamefont {Streif}\ \emph {et~al.}(2019)\citenamefont {Streif},
  \citenamefont {Neukart},\ and\ \citenamefont {Leib}}]{streif2019solving}%
  \BibitemOpen
  \bibfield  {author} {\bibinfo {author} {\bibfnamefont {M.}~\bibnamefont
  {Streif}}, \bibinfo {author} {\bibfnamefont {F.}~\bibnamefont {Neukart}},\
  and\ \bibinfo {author} {\bibfnamefont {M.}~\bibnamefont {Leib}},\ }\bibfield
  {title} {\bibinfo {title} {Solving quantum chemistry problems with a d-wave
  quantum annealer},\ }in\ \href@noop {} {\emph {\bibinfo {booktitle} {Quantum
  Technology and Optimization Problems: First International Workshop, QTOP
  2019, Munich, Germany, March 18, 2019, Proceedings 1}}}\ (\bibinfo
  {organization} {Springer},\ \bibinfo {year} {2019})\ pp.\ \bibinfo {pages}
  {111--122}\BibitemShut {NoStop}%
\bibitem [{\citenamefont {Seki}\ \emph {et~al.}(2021)\citenamefont {Seki},
  \citenamefont {Matsuzaki},\ and\ \citenamefont {Kawabata}}]{seki2021excited}%
  \BibitemOpen
  \bibfield  {author} {\bibinfo {author} {\bibfnamefont {Y.}~\bibnamefont
  {Seki}}, \bibinfo {author} {\bibfnamefont {Y.}~\bibnamefont {Matsuzaki}},\
  and\ \bibinfo {author} {\bibfnamefont {S.}~\bibnamefont {Kawabata}},\
  }\bibfield  {title} {\bibinfo {title} {Excited state search using quantum
  annealing},\ }\href@noop {} {\bibfield  {journal} {\bibinfo  {journal}
  {Journal of the Physical Society of Japan}\ }\textbf {\bibinfo {volume}
  {90}},\ \bibinfo {pages} {054002} (\bibinfo {year} {2021})}\BibitemShut
  {NoStop}%
\bibitem [{\citenamefont {Ura}\ \emph {et~al.}(2022)\citenamefont {Ura},
  \citenamefont {Imoto}, \citenamefont {Nikuni}, \citenamefont {Kawabata},\
  and\ \citenamefont {Matsuzaki}}]{ura2022analysis}%
  \BibitemOpen
  \bibfield  {author} {\bibinfo {author} {\bibfnamefont {K.}~\bibnamefont
  {Ura}}, \bibinfo {author} {\bibfnamefont {T.}~\bibnamefont {Imoto}}, \bibinfo
  {author} {\bibfnamefont {T.}~\bibnamefont {Nikuni}}, \bibinfo {author}
  {\bibfnamefont {S.}~\bibnamefont {Kawabata}},\ and\ \bibinfo {author}
  {\bibfnamefont {Y.}~\bibnamefont {Matsuzaki}},\ }\bibfield  {title} {\bibinfo
  {title} {Analysis of the shortest vector problems with the quantum annealing
  to search the excited states},\ }\href@noop {} {\bibfield  {journal}
  {\bibinfo  {journal} {arXiv preprint arXiv:2209.03721}\ } (\bibinfo {year}
  {2022})}\BibitemShut {NoStop}%
\bibitem [{\citenamefont {Serrano-Andr{\'e}s}\ and\ \citenamefont
  {Merch{\'a}n}(2005)}]{serrano2005quantum}%
  \BibitemOpen
  \bibfield  {author} {\bibinfo {author} {\bibfnamefont {L.}~\bibnamefont
  {Serrano-Andr{\'e}s}}\ and\ \bibinfo {author} {\bibfnamefont
  {M.}~\bibnamefont {Merch{\'a}n}},\ }\bibfield  {title} {\bibinfo {title}
  {Quantum chemistry of the excited state: 2005 overview},\ }\href@noop {}
  {\bibfield  {journal} {\bibinfo  {journal} {Journal of Molecular Structure:
  THEOCHEM}\ }\textbf {\bibinfo {volume} {729}},\ \bibinfo {pages} {99}
  (\bibinfo {year} {2005})}\BibitemShut {NoStop}%
\bibitem [{\citenamefont {Lacroix}\ \emph {et~al.}(2011)\citenamefont
  {Lacroix}, \citenamefont {Mendels},\ and\ \citenamefont
  {Mila}}]{lacroix2011introduction}%
  \BibitemOpen
  \bibfield  {author} {\bibinfo {author} {\bibfnamefont {C.}~\bibnamefont
  {Lacroix}}, \bibinfo {author} {\bibfnamefont {P.}~\bibnamefont {Mendels}},\
  and\ \bibinfo {author} {\bibfnamefont {F.}~\bibnamefont {Mila}},\ }\href@noop
  {} {\emph {\bibinfo {title} {Introduction to frustrated magnetism: materials,
  experiments, theory}}},\ Vol.\ \bibinfo {volume} {164}\ (\bibinfo
  {publisher} {Springer Science \& Business Media},\ \bibinfo {year}
  {2011})\BibitemShut {NoStop}%
\bibitem [{\citenamefont {Tokura}\ and\ \citenamefont
  {Kanazawa}(2020)}]{tokura2020magnetic}%
  \BibitemOpen
  \bibfield  {author} {\bibinfo {author} {\bibfnamefont {Y.}~\bibnamefont
  {Tokura}}\ and\ \bibinfo {author} {\bibfnamefont {N.}~\bibnamefont
  {Kanazawa}},\ }\bibfield  {title} {\bibinfo {title} {Magnetic skyrmion
  materials},\ }\href@noop {} {\bibfield  {journal} {\bibinfo  {journal}
  {Chemical Reviews}\ }\textbf {\bibinfo {volume} {121}},\ \bibinfo {pages}
  {2857} (\bibinfo {year} {2020})}\BibitemShut {NoStop}%
\bibitem [{\citenamefont {Advani}\ \emph {et~al.}(2020)\citenamefont {Advani},
  \citenamefont {Saxe},\ and\ \citenamefont {Sompolinsky}}]{advani2020high}%
  \BibitemOpen
  \bibfield  {author} {\bibinfo {author} {\bibfnamefont {M.~S.}\ \bibnamefont
  {Advani}}, \bibinfo {author} {\bibfnamefont {A.~M.}\ \bibnamefont {Saxe}},\
  and\ \bibinfo {author} {\bibfnamefont {H.}~\bibnamefont {Sompolinsky}},\
  }\bibfield  {title} {\bibinfo {title} {High-dimensional dynamics of
  generalization error in neural networks},\ }\href@noop {} {\bibfield
  {journal} {\bibinfo  {journal} {Neural Networks}\ }\textbf {\bibinfo {volume}
  {132}},\ \bibinfo {pages} {428} (\bibinfo {year} {2020})}\BibitemShut
  {NoStop}%
\bibitem [{\citenamefont {Imoto}\ \emph {et~al.}(2023)\citenamefont {Imoto},
  \citenamefont {Susa}, \citenamefont {Kadowaki}, \citenamefont {Miyazaki},\
  and\ \citenamefont {Matsuzaki}}]{imoto2023measurement}%
  \BibitemOpen
  \bibfield  {author} {\bibinfo {author} {\bibfnamefont {T.}~\bibnamefont
  {Imoto}}, \bibinfo {author} {\bibfnamefont {Y.}~\bibnamefont {Susa}},
  \bibinfo {author} {\bibfnamefont {T.}~\bibnamefont {Kadowaki}}, \bibinfo
  {author} {\bibfnamefont {R.}~\bibnamefont {Miyazaki}},\ and\ \bibinfo
  {author} {\bibfnamefont {Y.}~\bibnamefont {Matsuzaki}},\ }\bibfield  {title}
  {\bibinfo {title} {Measurement of the energy relaxation time of quantum
  states in quantum annealing with a d-wave machine},\ }\href@noop {}
  {\bibfield  {journal} {\bibinfo  {journal} {arXiv preprint arXiv:2302.10486}\
  } (\bibinfo {year} {2023})}\BibitemShut {NoStop}%
\bibitem [{\citenamefont {Seki}\ and\ \citenamefont
  {Nishimori}(2012)}]{seki2012quantum}%
  \BibitemOpen
  \bibfield  {author} {\bibinfo {author} {\bibfnamefont {Y.}~\bibnamefont
  {Seki}}\ and\ \bibinfo {author} {\bibfnamefont {H.}~\bibnamefont
  {Nishimori}},\ }\bibfield  {title} {\bibinfo {title} {Quantum annealing with
  antiferromagnetic fluctuations},\ }\href@noop {} {\bibfield  {journal}
  {\bibinfo  {journal} {Physical Review E}\ }\textbf {\bibinfo {volume} {85}},\
  \bibinfo {pages} {051112} (\bibinfo {year} {2012})}\BibitemShut {NoStop}%
\bibitem [{\citenamefont {Seki}\ and\ \citenamefont
  {Nishimori}(2015)}]{seki2015quantum}%
  \BibitemOpen
  \bibfield  {author} {\bibinfo {author} {\bibfnamefont {Y.}~\bibnamefont
  {Seki}}\ and\ \bibinfo {author} {\bibfnamefont {H.}~\bibnamefont
  {Nishimori}},\ }\bibfield  {title} {\bibinfo {title} {Quantum annealing with
  antiferromagnetic transverse interactions for the hopfield model},\
  }\href@noop {} {\bibfield  {journal} {\bibinfo  {journal} {Journal of Physics
  A: Mathematical and Theoretical}\ }\textbf {\bibinfo {volume} {48}},\
  \bibinfo {pages} {335301} (\bibinfo {year} {2015})}\BibitemShut {NoStop}%
\bibitem [{\citenamefont {Susa}\ \emph {et~al.}(2022)\citenamefont {Susa},
  \citenamefont {Imoto},\ and\ \citenamefont
  {Matsuzaki}}]{susa2022nonstoquastic}%
  \BibitemOpen
  \bibfield  {author} {\bibinfo {author} {\bibfnamefont {Y.}~\bibnamefont
  {Susa}}, \bibinfo {author} {\bibfnamefont {T.}~\bibnamefont {Imoto}},\ and\
  \bibinfo {author} {\bibfnamefont {Y.}~\bibnamefont {Matsuzaki}},\ }\bibfield
  {title} {\bibinfo {title} {Nonstoquastic catalyst for bifurcation-based
  quantum annealing of ferromagnetic $ p $-spin model},\ }\href@noop {}
  {\bibfield  {journal} {\bibinfo  {journal} {arXiv preprint arXiv:2209.01737}\
  } (\bibinfo {year} {2022})}\BibitemShut {NoStop}%
\bibitem [{\citenamefont {Susa}\ \emph {et~al.}(2018)\citenamefont {Susa},
  \citenamefont {Yamashiro}, \citenamefont {Yamamoto},\ and\ \citenamefont
  {Nishimori}}]{susa2018exponential}%
  \BibitemOpen
  \bibfield  {author} {\bibinfo {author} {\bibfnamefont {Y.}~\bibnamefont
  {Susa}}, \bibinfo {author} {\bibfnamefont {Y.}~\bibnamefont {Yamashiro}},
  \bibinfo {author} {\bibfnamefont {M.}~\bibnamefont {Yamamoto}},\ and\
  \bibinfo {author} {\bibfnamefont {H.}~\bibnamefont {Nishimori}},\ }\bibfield
  {title} {\bibinfo {title} {Exponential speedup of quantum annealing by
  inhomogeneous driving of the transverse field},\ }\href@noop {} {\bibfield
  {journal} {\bibinfo  {journal} {Journal of the Physical Society of Japan}\
  }\textbf {\bibinfo {volume} {87}},\ \bibinfo {pages} {023002} (\bibinfo
  {year} {2018})}\BibitemShut {NoStop}%
\bibitem [{\citenamefont {Pudenz}\ \emph {et~al.}(2014)\citenamefont {Pudenz},
  \citenamefont {Albash},\ and\ \citenamefont {Lidar}}]{pudenz2014error}%
  \BibitemOpen
  \bibfield  {author} {\bibinfo {author} {\bibfnamefont {K.~L.}\ \bibnamefont
  {Pudenz}}, \bibinfo {author} {\bibfnamefont {T.}~\bibnamefont {Albash}},\
  and\ \bibinfo {author} {\bibfnamefont {D.~A.}\ \bibnamefont {Lidar}},\
  }\bibfield  {title} {\bibinfo {title} {Error-corrected quantum annealing with
  hundreds of qubits},\ }\href@noop {} {\bibfield  {journal} {\bibinfo
  {journal} {Nature communications}\ }\textbf {\bibinfo {volume} {5}},\
  \bibinfo {pages} {3243} (\bibinfo {year} {2014})}\BibitemShut {NoStop}%
\bibitem [{\citenamefont {Pudenz}\ \emph {et~al.}(2015)\citenamefont {Pudenz},
  \citenamefont {Albash},\ and\ \citenamefont {Lidar}}]{pudenz2015quantum}%
  \BibitemOpen
  \bibfield  {author} {\bibinfo {author} {\bibfnamefont {K.~L.}\ \bibnamefont
  {Pudenz}}, \bibinfo {author} {\bibfnamefont {T.}~\bibnamefont {Albash}},\
  and\ \bibinfo {author} {\bibfnamefont {D.~A.}\ \bibnamefont {Lidar}},\
  }\bibfield  {title} {\bibinfo {title} {Quantum annealing correction for
  random ising problems},\ }\href@noop {} {\bibfield  {journal} {\bibinfo
  {journal} {Physical Review A}\ }\textbf {\bibinfo {volume} {91}},\ \bibinfo
  {pages} {042302} (\bibinfo {year} {2015})}\BibitemShut {NoStop}%
\bibitem [{\citenamefont {Imoto}\ \emph
  {et~al.}(2022{\natexlab{a}})\citenamefont {Imoto}, \citenamefont {Seki},
  \citenamefont {Matsuzaki},\ and\ \citenamefont
  {Kawabata}}]{imoto2022guaranteed}%
  \BibitemOpen
  \bibfield  {author} {\bibinfo {author} {\bibfnamefont {T.}~\bibnamefont
  {Imoto}}, \bibinfo {author} {\bibfnamefont {Y.}~\bibnamefont {Seki}},
  \bibinfo {author} {\bibfnamefont {Y.}~\bibnamefont {Matsuzaki}},\ and\
  \bibinfo {author} {\bibfnamefont {S.}~\bibnamefont {Kawabata}},\ }\bibfield
  {title} {\bibinfo {title} {Guaranteed-accuracy quantum annealing},\
  }\href@noop {} {\bibfield  {journal} {\bibinfo  {journal} {Physical Review
  A}\ }\textbf {\bibinfo {volume} {106}},\ \bibinfo {pages} {042615} (\bibinfo
  {year} {2022}{\natexlab{a}})}\BibitemShut {NoStop}%
\bibitem [{\citenamefont {Lidar}\ \emph {et~al.}(1998)\citenamefont {Lidar},
  \citenamefont {Chuang},\ and\ \citenamefont {Whaley}}]{lidar1998decoherence}%
  \BibitemOpen
  \bibfield  {author} {\bibinfo {author} {\bibfnamefont {D.~A.}\ \bibnamefont
  {Lidar}}, \bibinfo {author} {\bibfnamefont {I.~L.}\ \bibnamefont {Chuang}},\
  and\ \bibinfo {author} {\bibfnamefont {K.~B.}\ \bibnamefont {Whaley}},\
  }\bibfield  {title} {\bibinfo {title} {Decoherence-free subspaces for quantum
  computation},\ }\href@noop {} {\bibfield  {journal} {\bibinfo  {journal}
  {Physical Review Letters}\ }\textbf {\bibinfo {volume} {81}},\ \bibinfo
  {pages} {2594} (\bibinfo {year} {1998})}\BibitemShut {NoStop}%
\bibitem [{\citenamefont {Teplukhin}\ \emph {et~al.}(2019)\citenamefont
  {Teplukhin}, \citenamefont {Kendrick},\ and\ \citenamefont
  {Babikov}}]{teplukhin2019calculation}%
  \BibitemOpen
  \bibfield  {author} {\bibinfo {author} {\bibfnamefont {A.}~\bibnamefont
  {Teplukhin}}, \bibinfo {author} {\bibfnamefont {B.~K.}\ \bibnamefont
  {Kendrick}},\ and\ \bibinfo {author} {\bibfnamefont {D.}~\bibnamefont
  {Babikov}},\ }\bibfield  {title} {\bibinfo {title} {Calculation of molecular
  vibrational spectra on a quantum annealer},\ }\href@noop {} {\bibfield
  {journal} {\bibinfo  {journal} {Journal of chemical theory and computation}\
  }\textbf {\bibinfo {volume} {15}},\ \bibinfo {pages} {4555} (\bibinfo {year}
  {2019})}\BibitemShut {NoStop}%
\bibitem [{\citenamefont {Teplukhin}\ \emph {et~al.}(2020)\citenamefont
  {Teplukhin}, \citenamefont {Kendrick}, \citenamefont {Tretiak},\ and\
  \citenamefont {Dub}}]{teplukhin2020electronic}%
  \BibitemOpen
  \bibfield  {author} {\bibinfo {author} {\bibfnamefont {A.}~\bibnamefont
  {Teplukhin}}, \bibinfo {author} {\bibfnamefont {B.~K.}\ \bibnamefont
  {Kendrick}}, \bibinfo {author} {\bibfnamefont {S.}~\bibnamefont {Tretiak}},\
  and\ \bibinfo {author} {\bibfnamefont {P.~A.}\ \bibnamefont {Dub}},\
  }\bibfield  {title} {\bibinfo {title} {Electronic structure with direct
  diagonalization on a d-wave quantum annealer},\ }\href@noop {} {\bibfield
  {journal} {\bibinfo  {journal} {Scientific reports}\ }\textbf {\bibinfo
  {volume} {10}},\ \bibinfo {pages} {20753} (\bibinfo {year}
  {2020})}\BibitemShut {NoStop}%
\bibitem [{\citenamefont {Imoto}\ and\ \citenamefont
  {Matsuzaki}(2022)}]{imoto2022catastrophic}%
  \BibitemOpen
  \bibfield  {author} {\bibinfo {author} {\bibfnamefont {T.}~\bibnamefont
  {Imoto}}\ and\ \bibinfo {author} {\bibfnamefont {Y.}~\bibnamefont
  {Matsuzaki}},\ }\bibfield  {title} {\bibinfo {title} {Catastrophic failure of
  quantum annealing owing to non-stoquastic hamiltonian and its avoidance by
  decoherence},\ }\href@noop {} {\bibfield  {journal} {\bibinfo  {journal}
  {arXiv preprint arXiv:2209.10983}\ } (\bibinfo {year} {2022})}\BibitemShut
  {NoStop}%
\bibitem [{\citenamefont {Shammah}\ \emph {et~al.}(2018)\citenamefont
  {Shammah}, \citenamefont {Ahmed}, \citenamefont {Lambert}, \citenamefont
  {De~Liberato},\ and\ \citenamefont {Nori}}]{shammah2018open}%
  \BibitemOpen
  \bibfield  {author} {\bibinfo {author} {\bibfnamefont {N.}~\bibnamefont
  {Shammah}}, \bibinfo {author} {\bibfnamefont {S.}~\bibnamefont {Ahmed}},
  \bibinfo {author} {\bibfnamefont {N.}~\bibnamefont {Lambert}}, \bibinfo
  {author} {\bibfnamefont {S.}~\bibnamefont {De~Liberato}},\ and\ \bibinfo
  {author} {\bibfnamefont {F.}~\bibnamefont {Nori}},\ }\bibfield  {title}
  {\bibinfo {title} {Open quantum systems with local and collective incoherent
  processes: Efficient numerical simulations using permutational invariance},\
  }\href@noop {} {\bibfield  {journal} {\bibinfo  {journal} {Physical Review
  A}\ }\textbf {\bibinfo {volume} {98}},\ \bibinfo {pages} {063815} (\bibinfo
  {year} {2018})}\BibitemShut {NoStop}%
\bibitem [{\citenamefont {Imoto}\ \emph
  {et~al.}(2022{\natexlab{b}})\citenamefont {Imoto}, \citenamefont {Seki},\
  and\ \citenamefont {Matsuzaki}}]{imoto2022quantum}%
  \BibitemOpen
  \bibfield  {author} {\bibinfo {author} {\bibfnamefont {T.}~\bibnamefont
  {Imoto}}, \bibinfo {author} {\bibfnamefont {Y.}~\bibnamefont {Seki}},\ and\
  \bibinfo {author} {\bibfnamefont {Y.}~\bibnamefont {Matsuzaki}},\ }\bibfield
  {title} {\bibinfo {title} {Quantum annealing with symmetric subspaces},\
  }\href@noop {} {\bibfield  {journal} {\bibinfo  {journal} {arXiv preprint
  arXiv:2209.09575}\ } (\bibinfo {year} {2022}{\natexlab{b}})}\BibitemShut
  {NoStop}%
\end{thebibliography}%

\appendix

\section{How to select the initial longitudinal magnetic field}\label{sec:make_hl}

To perform our excited state search, we need to start from an Ising Hamiltonian with the longitudinal magnetic field where we can efficiently find the ground state.
Also, we need to resolve the degeneracy of the target excited state.
We consider the following longitudinal magnetic field 
\begin{align}
    H_{L}=\sum_{j=1}^{L}h_{j}\hat{\sigma}^{(z)}_{j}.\label{longappendix}
\end{align}
We can resolve the degeneracy up to ($2^{k}-1)$-th excited states by choosing the coefficients of the longitudinal magnetic field as follows
\begin{align}
    h_{l}=
       \begin{cases}
      2^{l-1}h\ \  (l\leq k)\\
      2^{k-1}h\ \  (k<l).\label{eq:h_def}
   \end{cases}
\end{align}
We suppose the initial longitudinal magnetic field satisfies the Eq. (\ref{eq:h_def}) throughout this appendix.
By setting a large $h$, we can efficiently find a ground state of the Ising Hamiltonian with the longitudnal field, because the Ising interaction can be negligible compared with the energy splitting due to the longitudinal field.
The problem Ising Hamiltonian is given as 
\begin{align}
    H_{P}&=\sum_{i>j}J_{ij}\hat{\sigma}^{(z)}_{i}\hat{\sigma}^{(z)}_{j}
\end{align}
where the maximum (minimum) energy of this Hamiltonian is $\sum_{i>j}|J_{ij}|$ ($-\sum_{i>j}|J_{ij}|$).
On the other hand, if we consider the longitudinal magnetic field Hamiltonian in Eq. \eqref{longappendix}, the energy gap between the $j$-th and $j+1$-th excited states up to $(2^{k}-1)$-th excited states is $2h$.
Therefore, if
\begin{align}
\sum_{i>j}|J_{ij}|<h\label{eq;cond_Jh}
\end{align}
is satisfied, the target excited state of the Hamiltonian $H_P+H_L$ is the same that of $H_P$. 
This means that we can efficiently find the excited state.

In this approach, we need to apply larger longitudinal fields as we increase the size of the problem Hamiltonian. On the other hand,
if the transverse field and Ising Hamiltonian scheduling can be designed independently on a D-wave machine, our method can be applied to large-scale problems without such large longitudinal fields\cite{seki2021excited}.
Implementation of this function expands the potential applications of the D-wave machine.
We discuss this point in the Appendix \ref{futureappendix}.

\section{Energy diagram with and without the transverse magnetic field}

\begin{figure}[h!]
    \centering
    \includegraphics[width=70mm]{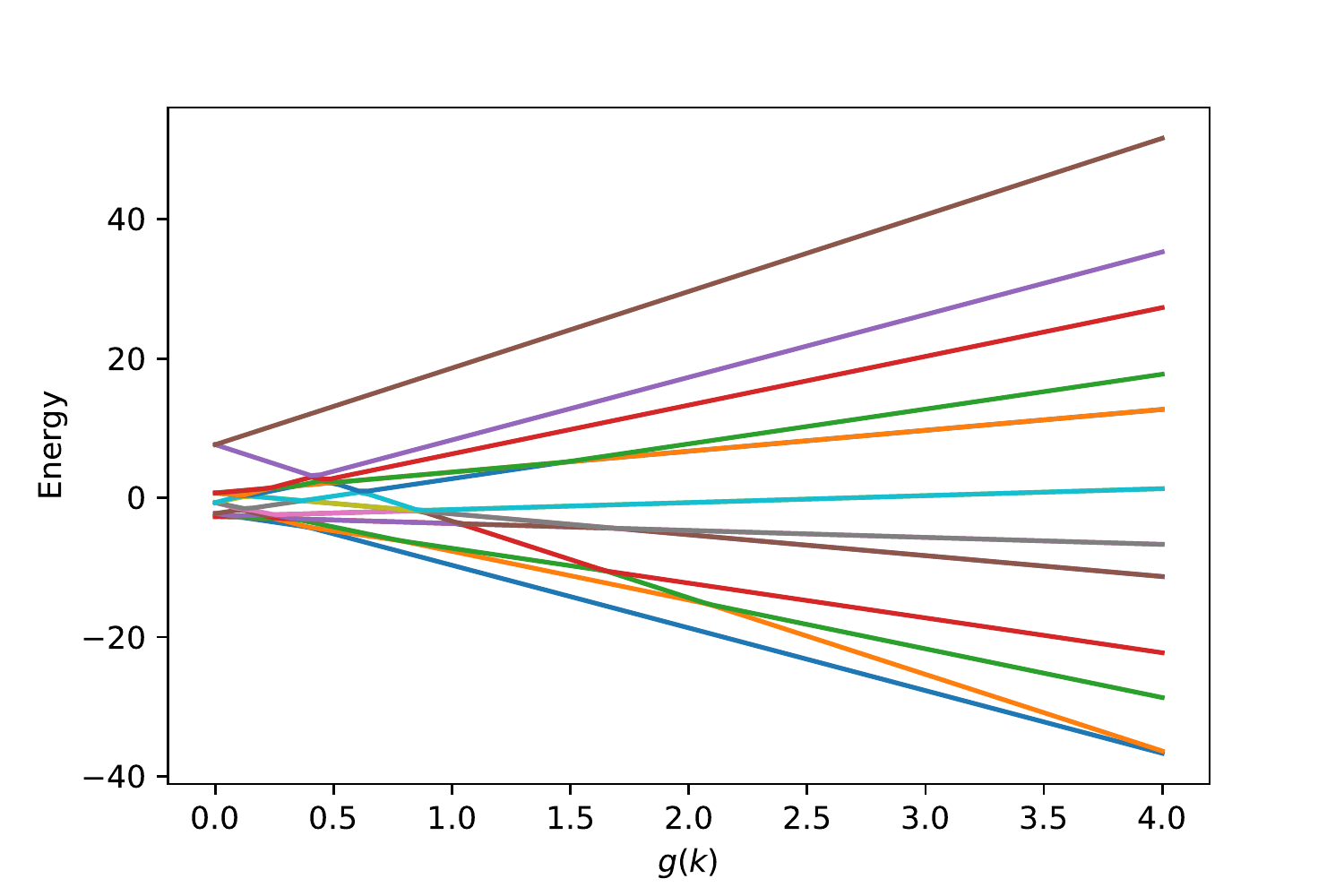
    }
    \caption{The energy diagram of $g(k)H_{L}+H_{P}$ against $g(k)$.  }\label{fig:energy_diagram_without_tf_svp}
\end{figure}

\begin{figure}[h!]
    \centering
    \includegraphics[width=80mm]{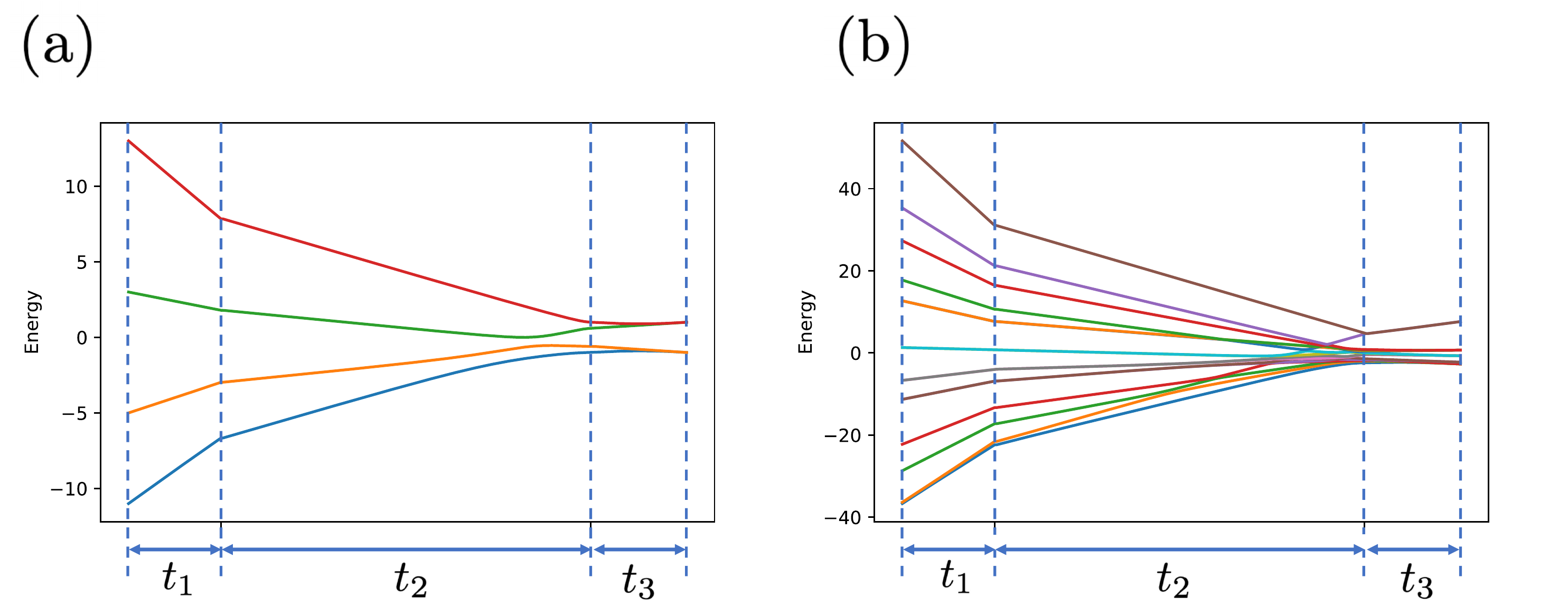
    }
    \caption{(a) The energy diagram during our excited state search of the 2-qubit model with the transverse magnetic field.
    (b) The energy diagram during our excited state search of the SVP with the transverse magnetic field.
    }\label{fig:energy_diagram_with_tf_2qubit_and_svp}
\end{figure}

In this section, we show the energy diagram during our method with and without the transverse magnetic field.
We plot the energy diagram of $g(k)H_{L}+H_{P}$ for the SVP against $g(k)$ in FIG.\ref{fig:energy_diagram_without_tf_svp}.
There are level crossings for the two-qubit model (see FIG.\ref{fig:D-wave_result_two_qubit} (a)).
In addition, we describe the energy diagram during our method for the two-qubit model and the SVP in FIG. \ref{fig:energy_diagram_with_tf_2qubit_and_svp}.
Unlike the case without transverse magnetic fields, there are avoided level crossing
(see FIG.\ref{fig:D-wave_result_two_qubit} (a) and FIG.\ref{fig:energy_diagram_without_tf_svp}).

\section{Transformation of the SVP into Ising Hamiltonian}\label{sec:mapping}

\begin{figure}[h!]
    \centering
    \includegraphics[width=70mm]{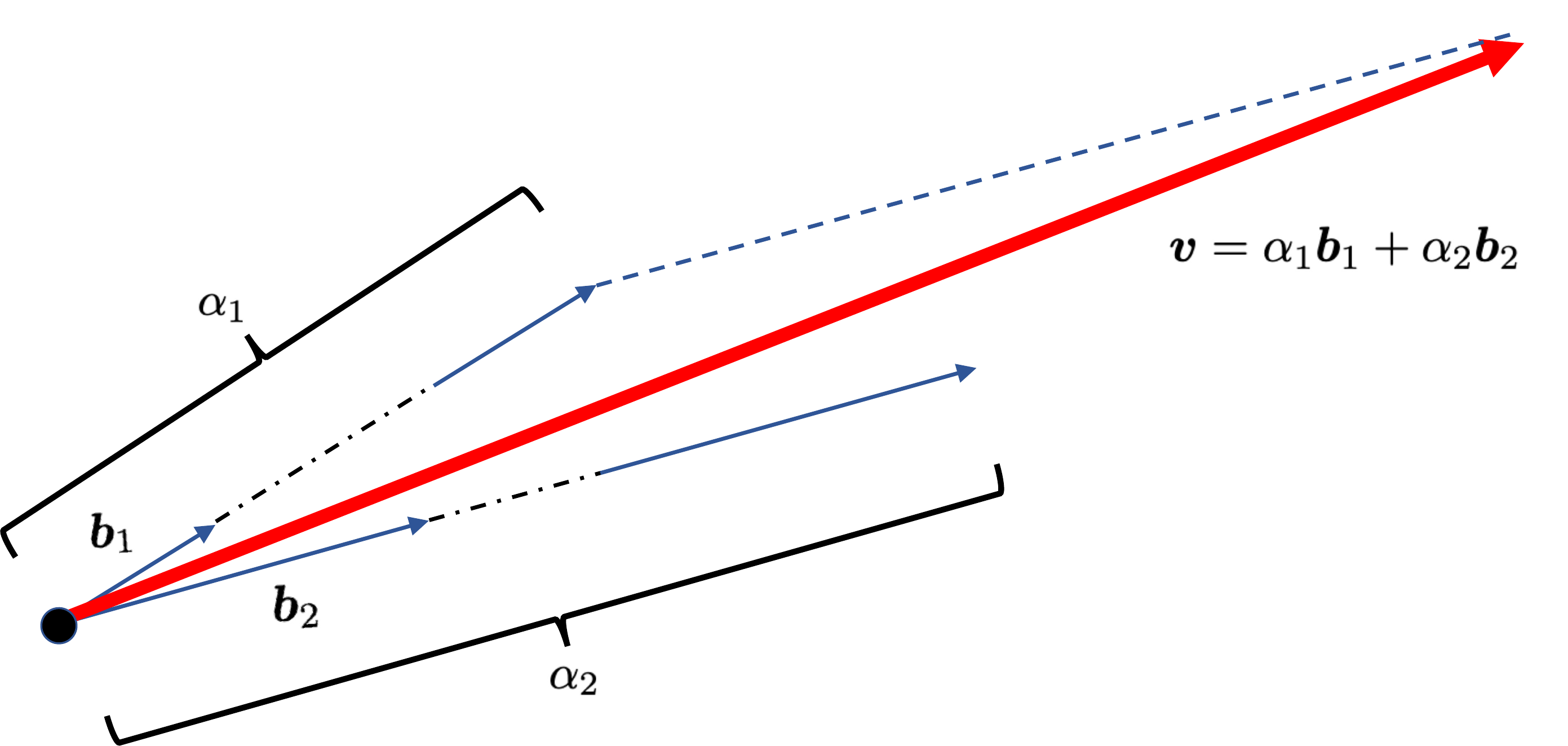
    }
    \caption{
    Illustration of a lattice vector for $L=2$. This vector is represented by $\bm{v}=\alpha_{1}\bm{b}_{1}+\alpha_{2}\bm{b}_{2}$ where $\{\bm{b}_{j}\}_{j=1}^{2}$ 
    is a set of basis and $\{\bm{\alpha}_{j}\}_{j=1}^{2}$ is a set of integer.
    SVP is the problem of finding the $\alpha_{1}$ and $\alpha_{2}$ to minimize the norm of $\bm{v}$ except the origin.
    }\label{fig:svp_fig}
\end{figure}

In this section, we review the SVP, and explain how to map the SVP onto the Ising Hamiltonian \cite{joseph2021two}.
Let us define a set of linearly independent basis vectors  $\{\bm{b}_{j}\}_{j=1}^{L}$ where the dimension is $L$. 
Also, we define a set of integers $\{\alpha_{j}\}_{j=1}^{L}$, which represents the coefficients of the lattice basis.
We consider an $L$ dimensional lattice vector $\bm{v}$ as follows.
\begin{align}
    \bm{v}=\sum_{j=1}^{L}\alpha_{j}\bm{b}_{j}.
\end{align}
The aim of the SVP is to find a vector with the smallest norm, except the zero vector(see Fig \ref{fig:svp_fig}).

When we adopt the so-called Hamming encoding, the SVP can be mapped onto the Ising Hamiltonian as follows.
The square of the norm of the vector $\bm{v}$ is given by
\begin{align}
||\bm{v}||^{2}&=\sum_{i,j=1}^{L}c_{i}c_{j}\bm{b}_{i}\cdot\bm{b}_{j}\\
    &=\sum_{i,j=1}^{L}c_{i}c_{j}\bm{G}_{i,j}\label{eq:svp_vec}
\end{align}
where $\mathbf{G}_{i,j}=\bm{b}_{i}\cdot\bm{b}_{j}$ is the element of the Gram matrix of the lattice basis vectors.
Let us consider a search for the solution of SVP in the range $-k\leq c_{j}\leq k,\ \ j=1,\cdots, L$.
We replace the coefficient of the basis vector with the following operator
\begin{align}
    \hat{Q}^{(i)}=\sum_{p=1}^{2k}\frac{\hat{\sigma}_{p,j}^{(z)}}{2},\ \ \ j=1,\cdots L\label{eq:Q_def}
\end{align}
where $\hat{Q}^{(i)}$ is a diagonal matrix.
The diagonal elements of this matrix correspond to the value of the coefficient of the lattice basis.
Using this matrix, we represent the square of the lattice vector (\ref{eq:svp_vec}) as follows.

\begin{align}
    \hat{H}^{(SVP)}=\sum_{i,j=1}^{L}\bm{G}_{i,j}\hat{Q}^{(i)}\hat{Q}^{(j)}.
\end{align}
This Hamiltonian consists of $2kL$ qubits.
If we consider a subspace spanned by Dicke basis with the maximum total angular momentum, the first excited state of this Hamiltonian is the solution of the SVP, while the ground state is the zero vector \cite{ura2022analysis}. 
On the other hand, if we consider a full Hilbert space, the ground states are degenerate, and the number of degenerate ground states is $({}_{2k}C_k)^{L}$, and the first excited state is the solution.

\section{Application to solve a problem for large input sizes}\label{futureappendix}

Let us discuss how to solve a problem for large input sizes by using the excited state search.
We must prepare a substantially large longitudinal magnetic field to satisfy the condition (\ref{eq;cond_Jh}), when our method can be applied to large-scale problems.
However, if the transverse field and Ising Hamiltonian scheduling can be designed independently on a D-wave machine, the left side hand of the inequality (\ref{eq;cond_Jh}) can be set to zero at $t=0$ when we prepare the initial excited state, and we can gradually increase the value of the Ising interaction.
Thus, we don't need a substantially large longitudinal magnetic field to apply our method.
We believe this function will be realized in the near future.

In addition, to solve the SVP by using the full Hilbert space, we have to prepare the $(({}_{2k}C_k)^{L}+1)$-th excited state in our current method as we discussed in the Appendix \ref{sec:mapping}.
In this case, we need to resolve an exponentially large number of degeneracies, and the energy gap between the target eigenstate and another eigenstate could be exponentially small.

To avoid the difficulty to resolve an exponentially large number of degeneracies, we can adopt a strategy to use a subspace spanned by the Dicke states, as proposed in \cite{ura2022analysis}.
We adopt the uniform longitudinal magnetic field as the initial driving Hamiltonian.
When we consider the Dicke basis, the eigenstates of $\hat{Q}^{(i)}$ and longitudinal magnetic field (defined by (\ref{eq:Q_def})) are spanned by 
\begin{align}
\ket{W_{r}}=\sum_{p_{1},\cdots,p_{r}=1}^{2k}\hat{\sigma}_{p_{1}}^{(x)}\cdots\hat{\sigma}_{p_{r}}^{(x)}\bigotimes_{j=1}^{2k}\ket{\downarrow}_{j},\ \ r=0,1,2,\cdots,2k.
\end{align}
which is called the Dicke state.
In this case, the ground state (the first excited state) of the uniform longitudinal magnetic field is represented by $\ket{W_{r=0}}$($\ket{W_{r=1}}$).
If we use the subspace spanned by the Dicke states, the first excited state of the problem Hamiltonian is the solution of the SVP. Such an approach to use the subspace was proposed in \cite{shammah2018open, ura2022analysis, imoto2022quantum} for the purpose to suppress the non-adiabatic transitions.
However, the potential problem is the difficulty to prepare the Dicke state because the state is entangled, and we may not be able to prepare the entangled state as the initial state by using the D-wave machine in the near future.
Thus, in order to solve SVP by our excited state search without preparing Dicke states in D-wave, we adopt the following strategy.
We prepare the following separable states.
\begin{align}
    \rho= \frac{1}{2k}\sum _{p=1}^{2k} \hat{\sigma}_{p}^{(x)}(\bigotimes_{j=1}^{2k}\ket{\downarrow}_{j}\bra{\downarrow})\hat{\sigma}_{p}^{(x)},\ \ p=1,\cdots,2k.\label{eq:separate_init_state}
\end{align}
We perform our excited state search using this initial state (\ref{eq:separate_init_state}).
We have $\bra{W_{r}} \rho\ket{W_{r}}=1/r$. 
Starting from the initial state $\ket{W_{r}}$, we can obtain the solution with a unit success probability 
if an adiabatic condition is satisfied.
Therefore, even if we start from the initial state of $\rho$, the success probability of finding the solution is $1/r$ as long as the adiabatic condition is satisfied.
Of course, there is a possibility that it takes an exponentially large time to satisfy the adiabatic condition depending on the energy gap during QA. The detailed study of the energy gap during the excited state search to solve the SVP for a large size is left for future work.

\end{document}